**Physics-informed neural networks for multi-field visualization with single-color laser induced fluorescence**

**Short title:** Deep Neural Networks for Multi-Field Visualization

Nagahiro Ohashi[1], Leslie K. Hwang[2] and Beomjin Kwon[1]*

[1]School for Engineering of Matter, Transport and Energy, Arizona State University, Tempe, AZ 85287, USA

[2]School of Electrical, Computer and Energy Engineering, Electrical Engineering, Arizona State University, Tempe, AZ 85281, USA

*Corresponding author: kwon@asu.edu)

**Abstract**

Reconstructing fields from sparsely observed data is an ill-posed problem that arises in many engineering and science applications. Here, we investigate the use of physics-informed neural networks (PINNs) to reconstruct complete temperature, velocity and pressure fields from sparse and noisy experimental temperature data obtained through single-color laser-induced fluorescence (LIF). The PINNs are applied to the laminar mixed convection system, a complex but fundamentally important phenomenon characterized by the simultaneous presence of transient forced and natural convection behaviors. To enhance computation efficiency, this study also explores transfer learning (TL) as a mean of significantly reducing the time required for field reconstruction. Our findings demonstrate that PINNs are effective, capable of eliminating most experimental noise that does not conform to governing physics laws. Additionally, we show that the TL method achieves errors within 5% compared to the regular training scheme while reducing computation time by a factor of 9.9. We validate the PINN reconstruction results using non-simultaneous particle image velocimetry (PIV) and finite volume method (FVM) simulations. The reconstructed velocity fields from the PINN closely match those obtained from PIV. When using FVM data as a reference, the average temperature errors are below 1%, while the pressure and velocity errors are below 10%. This research provides insights into the feasibility of using PINNs for solving ill-posed problems with experimental data and highlights the potential of TL to enable near real-time field reconstruction.

**Introduction**

Laser-induced fluorescence (LIF) is a spectroscopic method used to measure the temperature fields in liquids by exciting the fluorescent dyes with a laser and measuring the emitted fluorescence [1,2]. The intensity of the emitted fluorescence, $I_1$ (W/m$^3$), is given as $I_1=I_0C_1\phi_1\varepsilon_1$, where $I_0$ is the incident laser flux (W/m$^2$), C is the dye concentration (kg/m$^3$), $\phi$ is the quantum efficiency, $\varepsilon$ is an absorption coefficient (m$^2$/kg), and the subscript 1 denotes a primary fluorescent dye. For a typical LIF dye, $\phi$ is temperature-dependent, while $\varepsilon$ is temperature-independent. Thus, if C and $I_0$ are kept constant, it is possible to relate $I_1$ to the local fluid temperature. However, $I_0$ may not be uniform across the fluid volume due to various effects, such as refractions through entrained air bubbles and nonuniform temperature fields, both of which cause variations in the refractive index. To mitigate this problem, two-color LIF (2-CLIF) technique is commonly applied, which acquires $I_0$ field simultaneously during the LIF measurement using a second fluorescent dye. Typically, Rhodamine 110 is used for the purpose, that has a temperature-independent $\phi$ and an emission spectrum distinct from the temperature-sensitive dye. The fluorescence intensity of the second dye, $I_2$, is temperature-insensitive but sensitive to the $I_0$ field. By using both $I_1$ and $I_2$, the fluorescence signal ratio, $I_1/I_2=C_1\phi_1\varepsilon_1/C_2\phi_2\varepsilon_2$, is obtained, which is yet temperature-dependent, but independent of $I_0$, effectively correcting artifacts due to nonuniform $I_0$ [3–5].

The 2-CLIF has been successfully applied to various problems, such as temperature measurements in micro-scale heated channels and high aspect-ratio Rayleigh-Benard convection, achieving an accuracy of 1.5 K [3,6,7]. Additionally, this method has been used in stratified flows and steady-state microfluidic channels [8,9], with accuracies of 1.45 K and 0.3K, respectively. Lastly, research on heated turbulent jets injected into coflow has visually compared the differences between single-color LIF and 2-CLIF [10]. The 2-CLIF method demonstrated more accurate measurements of symmetric temperature distributions, with a temperature error of only 1℃, compared to a 4℃ error in the single-color LIF [10]. However, the 2-CLIF requires a more complex optical setup compared to a single-dye LIF system, including two separate cameras, two signal processing units, and additional optical components.

Furthermore, when both velocity and temperature fields need to be visualized for a comprehensive understanding of fluid flows, particle image velocimetry (PIV) can be conducted simultaneously with LIF. The simultaneous LIF-PIV has been demonstrated in some applications such as visualizing stratified flow and impinging jet [11]. In another demonstration, simultaneous 2-CLIF, PIV, and infrared thermometry were employed to study the flow and heat transfer in nucleate boiling phenomena [12]. Other research included measurements of flow velocity in a transient diesel spray [13], temperature and velocity of transient buoyant plume above a horizontal cylinder [14] and parallel plane jets [15]. Yet again, such combined use of visualization techniques requires significant experimental resources, including multiple cameras and data analysis software.

To enable the visualization of temperature and velocity fields without combining multiple hardware resources, we explore the use of artificial neural networks, specifically physics-informed neural networks (PINNs), to reconstruct and predict these fields using only a single-dye LIF system. PINNs are a class of neural networks designed to approximate the solutions of partial differential equations (PDEs) governing the physics of a system. Unlike traditional machine learning methods that solely rely on large datasets, PINNs use PDEs to constrain the domain space during the learning process, reducing the need for extensive data. Seminal work by Raissi et al. introduced PINNs for addressing forward and inverse nonlinear problems, using Burger and Schrodinger equations as examples [16]. PINNs have recently shown promise in solving inverse problems, such as reconstructing velocity, temperature or other flow variables based on partial or indirect measurements, which is a significant yet challenging task.

Reconstructing thermofluidic fields using PINNs has been demonstrated in several studies, though most have utilized numerically generated noiseless data. In an early demonstration, a two-dimensional (2D), quasi-steady natural convection flow around a cylinder was reconstructed using 10 data frames, with each frame containing 200 of both temperature and velocity data points [17]. Next, the reconstruction of a three-dimensional (3D) flame in transient turbulent combustion was achieved using temperature, velocity, and mass fraction data points acquired from direct numerical simulation, in which three PINN sub-models were used to learn each governing equation [18]. In another study, a 2D mixed convection within a square domain with a heated bottom was reconstructed [19]. Mixed convection involves complex transient phenomena where buoyancy driven flows contribute to the overall flow characteristic of a system. The PINN was trained only on temperature data obtained from computational fluid dynamics (CFD) simulations and reconstructed the unknown pressure and velocity fields. This approach was further extended to demonstrate experimental transient natural convection. The unsteady 3D temperature field in the flow over an espresso cup was captured using Tomographic Background Oriented Schlieren, and 400 frames of temperature data without boundary conditions were used to reconstruct the simultaneous velocity fields [19].

To improve the versatility of PINNs, it is also crucial to reduce the training time required for reconstructing unseen flows. In this study, we explore the transfer learning (TL) approach to expedite the training of PINNs on unfamiliar yet related flows. Previous research on transfer learning for PINNs (TLPINNs) has revealed that similar accuracies can be achieved with reduced training time and less data [20]. For instance, studies on simple linear ordinary and partial differential equations demonstrated that freezing certain hidden layers of a pre-trained PINN during the TL process allowed optimization to be confined to a single linear layer [21]. Another investigation applied TLPINNs to predict various structural loading scenarios using limited displacement information, achieving satisfactory results (within 5% errors) even with noisy and incomplete data [22]. Additionally, TLPINN was used for flow field reconstruction in vortex-induced vibration scenarios. A pre-trained PINN was retrained using ½, ¼, and ⅛ of the original dataset, resulting in about ½ (22.9 hr), ⅕ (10.3 hr), and 1/9 (5.4 hr) the original computation time, respectively. Although using less data led to higher error, all maximum relative errors remained within 4.9% [23].

Here, we reconstruct the temperature, velocity, and pressure fields of mixed convection flows using noisy temperature data obtained from a single-dye LIF system. To achieve this, a PINN is optimized and trained on a specific time frame of the flow. This trained PINN then serves as the basis for training additional PINNs to reconstruct unfamiliar flows at different time frames. By employing a TL strategy, the subsequent training of PINNs is significantly accelerated. The reconstruction results are indirectly validated by comparing them with CFD simulations and non-simultaneous PIV data for the same flow configuration.

**Results**

Experimental LIF data were collected for a rectangular channel heated from the bottom at a constant temperature of 32℃ using a dyed deionized water as working fluid. Figure 1 (a,b) shows the LIF data acquisition in the field of interest (FOI), which measured 32 mm in width and 20 mm in height. As the working fluid at 30℃ passed through, the laser excited the temperature-sensitive dye, which emitted fluorescence captured by a high-speed camera. A schematic showing the fluid loop and data loop can be found in the Appendix A. For PINN training, a dataset consisting of 100 time frames with 0.1-second intervals were sampled from the LIF data, referred to as the entire dataset. The time coordinate was normalized as a nondimensional time coordinate ($t^*$) ranging from 0 to 1.

Figure 1(c) depicts the PINN structure and training process, where non-dimensional spatial ($x^*$, $y^*$) and temporal ($t^*$) coordinates are used. We implement a multiple-scale PINN (MSPINN), which incorporates a scaling layer prior to the traditional PINN configuration - a fully connected neural network (FCNN) - as detailed in the Appendix C. The scaling layer adjusts the inputs before they are processed by the FCNN, which consists of 10 layers. Previous studies have demonstrated that multiple-scale problems in machine learning required additional architecture for better results [24,25], particularly in phenomena involving features across multiple spatial or temporal scales, such as stiff chemical kinetics and bending problems [26,27]. In channel flows, multiple scales arise due to the manifestation of boundary layers that feature different scales compared to the characteristic length of the flow. Similarly, in our problem, buoyancy-driven plumes can develop on several length and time scales, significantly impacting the overall flow dynamics.

Successful PINN training required a minimum amount of experimental data. For our PINN architecture, tests determined that 200 data points per $mm^2$ and a sampling period of 0.2 seconds were sufficient to reconstruct the temperature field within a 10% discrepancy compared to using the entire dataset, which is detailed in the Appendix E. These parameters were used for subsequent PINN training unless stated otherwise.

The optimization of the PINN was achieved by minimizing a composite loss function that contains constraints on boundary conditions, data points, and physics laws, refer to Appendix D for detailed equations for these loss functions. For the mixed convection channel flow with a small temperature variation (e.g., ΔT < 2K), the relevant thermofluidic physics were governed by the conservation of mass, the laminar Navier-Stokes equations with the Boussinesq approximation, and the conservation of energy. The nondimensionalized versions of these PDEs are detailed in the Appendix B and were used to compute the physics losses of the PINN solutions, which were the residuals of the PDEs.

The losses against the boundary conditions were also considered. During most LIF measurements, only limited boundary conditions were known, such as fluid velocity at the free surface and channel walls. Boundary conditions that required additional sensors, such as temperature and pressure, as well as velocity distributions across the channel inlet and outlet, were not used to guide the PINN search space. Points along the known boundary, where the boundary condition loss was calculated, were sampled uniformly in the $x^*$, $y^*$, and $t^*$ directions.

Data loss, which is the difference between LIF temperature measurements and the PINN solution, was evaluated at points randomly sampled throughout the FOI. Only temperature data was used to compute the data loss, with no additional field data. Special care is necessary when parsing the LIF data, since the fidelity of the data depends on calibration quality and local noise level. Data preprocessing ensures that the sampled temperatures are within the actual range (i.e., between 30℃ and 32℃ in our experiment). In addition to the data points, a portion of the collocation points, where the losses against PDEs are calculated, were randomly sampled. The distribution of points is shown in Fig. 1(c), along with a sample reconstruction result. To save computation memory, the coordinates of data points also served as those of collocation points, eliminating the generation of additional collocation points. Using 20,200 boundary points, 64,500 data points, 20,000 collocation points and 100,000 Adam epochs, the PINN training took about 12 hours and used around 40 GB of memory. Equal weights were used between PDE, boundary condition and data losses in the composite loss function unless otherwise specified.

Reconstruction of flow fields using LIF temperature data was performed on three datasets, denoted as LIF-1, LIF-2, and LIF-3, with 100 frames each. During the LIF measurements, the flow condition was transient mixed convection, characterized by a Reynolds number (Re) of 50 and a Richardson (Ri) number of 403. Under the same flow condition, these datasets were sampled, capturing unique flow features at different moments, as shown in the snapshots at four different t* for each dataset in Fig. 2. A cold-water stream appeared in the middle of FOI. Due to the diverging shape of the inlet port, the cold stream naturally flowed from top left to bottom right. Complex variations in flow fields and flow mixing constantly occurred due to the sudden generation of buoyancy driven flows, originating from the heated bottom surface. Since the dataset was captured after the flow field was developed, low-density hot fluid occupied the top portion of the channel. In the raw single-color LIF data, common noises were 1) narrow vertical lines, 2) large-scale uneven backgrounds, 3) discontinuous dotted regions, and 4) discontinuous temperature fields, which appeared mainly due to the non-uniform laser intensity distribution.

Figure 3 shows the reconstruction results at selected t*, representing the temperature, pressure, x-velocity (U*), y-velocity (V*) and velocity vector maps. The results shown in Fig. 3 correspond to two t* in Fig. 2 that exhibit the most significant flow field variations for each dataset. A video showing the full-frame reconstruction is available online [28]. For single-color LIF noise manifesting at higher frequencies (e.g., localized noises) than the underlying lower-frequency temperature field structures, the inherent preference of PINNs for low-frequency structures is leveraged to reconstruct fields from noisy temperature data. Generally, PINNs function as low-pass filters, allowing for the recovery of most of the original data. Visual comparison between Fig. 2 and 3 demonstrates the ability of PINNs to filter out the localized noises from LIF data, resulting in a cleaner temperature field output. This result suggests that the PINN effectively distinguishes between correct and noisy inputs during training. Although the noisy experimental data influences the search space, the PINN refines its solution by complying with the governing PDEs and boundary conditions. Data points that fail to satisfy the PDEs (i.e., noise) are not likely to be enforced due to potential increase in PDE loss. Consequently, the PINN prioritizes the enforcement of correct data points, as evidenced by the plots in the Appendix F, showing the training loss versus epoch, which indicates that the model converges under the noisy LIF data constraints.

Next, we explored the application of transfer learning (TL) to reduce the training time of PINN. Four datasets, referred to as TL-0 through TL-3, were sampled at different times under the same flow condition as the previous datasets. Specifically, TL-0 was used to train a baseline PINN, whose neural network parameters were then transferred to train additional PINNs based on TL-1 through TL-3. Figure 4 presents snapshots of raw LIF data at four different $t^*$ for each dataset. The four datasets exhibited similar yet distinct patterns of temperature and noise distributions.

New PINNs were trained using TL-1, TL-2 or TL-3, initialized with the parameters of the TL-0 model (denoted as TLPINNs). A significant advantage of the TL scheme is that it leverages a pre-trained model (i.e., TL-0 model) to improve the training efficiency of new models on related tasks, reducing the number of training epochs required. To optimize training duration, these PINNs were trained with different numbers of Adam epochs, as detailed in the Appendix G. The analysis shows that 1000 Adam epochs are sufficient, representing merely 1% of the epochs required to train a randomly initialized PINN (denoted as $PINN_0$), which requires 100,000 epochs to train to achieve high accuracy.

Figure 5 presents the reconstruction results of the TLPINNs compared to $PINN_0$, in which one frame of each dataset from Fig. 4 is reconstructed, exhibiting the most complex pattern. The metric used to evaluate the PINNs is the mean absolute error ($\epsilon$), defined as $\epsilon_F = \sum_0^N |(F_{PINN} - F_{ref})|/N$, where $|(F_{PINN} - F_{ref})|$ represents the absolute difference between a field variable predicted by the PINN (i.e., $F_{PINN}$) and the reference data (i.e., $F_{ref}$). Here, *N* is the number of points for error estimation, and the subscript *F* denotes the type of field. Comparing the $T^*$ from Fig. 4 and Fig. 5 demonstrates that the TLPINNs effectively eliminate noise and capture the major patterns. All $\epsilon$ values are within 5% when using $PINN_0$ as reference. The comparison with $PINN_0$ shows that TLPINNs produce similar reconstructed fields with slight variations. Note that, for the more complex flow in TL-3, the deviation from $PINN_0$ is greater. However, the general flow pattern, such as circulation in the bottom left corner, is captured in all reconstruction results. Comparing to $PINN_0$, the TL method was able to achieve approximately 9.9 times less computational time on average, further details are provided in the Appendix G.

To validate the robustness of the PINN reconstruction results, we conducted non-simultaneous PIV and FVM simulations for comparison with the PINN solutions. First, non-simultaneous PIV measurements of the velocity field allowed us to compare the velocity reconstruction from PINN against experimental data taken under identical flow conditions. Figure 6 presents three PIV snapshots (PIV-1, PIV-2, and PIV-3), that exhibit similar velocity fields to those reconstructed in Fig. 3 (LIF-3, $t^*=0.33$) and Fig. 5 (TL-1, $t^*=0.66$). The PIV data demonstrates velocity maps that closely match those reconstructed by PINN. Specifically, $U^*$ map has a dominant high velocity stream in the middle, with lower velocity at the top. $V^*$ map constantly changes as the flow develops, however, the pattern of higher $V^*$ in top right and lower $V^*$ in bottom right is consistent in both the PINN reconstruction and the PIV data. The development of the PIV field is further visualized in the Appendix H.

In addition to the PIV measurements, a 3D FVM employing the channel geometry shown in Fig. 1(a) was developed. The FVM generated a complete set of fields for three instances representing different FOI positions at Re=50 and a Ri=403, referred to as FVM-1, FVM-2, and FVM-3, as shown in Fig. 7. These datasets were chosen to study a variety of temperature field complexities. For reconstructing simple temperature fields (FVM-2 and FVM-3), equal weights between loss terms were used. However, the reconstruction for the complex temperature field (FVM-1) required increasing the data loss weight to 50 to obtain acceptable results. A video showing the full-frame reconstruction is available online [29].

The reconstruction of a complete set of fields based only on the FVM temperature field demonstrates that the accuracy of PINN outputs is influenced by the complexity of the temperature variation over time. Using the FVM data as a reference, $\epsilon_{T^*}$ that is averaged over all frames are below 1% for all datasets. For FVM-2 and FVM-3, the reconstruction produces time-averaged $\epsilon_{P^*}$, $\epsilon_{U^*}$, and $\epsilon_{V^*}$ below 10%, while FVM-1 results in time-averaged $\epsilon_{P^*}$, $\epsilon_{U^*}$, and $\epsilon_{V^*}$ of 16%, 14%, and 12%, respectively. The complexity of the temperature field variation is quantified by field-averaged temporal gradients (denoted as $\nabla_t T^*$), representing the magnitude of local field variation over time. FVM-1 exhibits a $\nabla_t T^*$ value approximately twice as high as those of FVM-2 and FVM-3 (e.g., $\nabla_t T^*_{FVM-1} = 0.0035$, $\nabla_t T^*_{FVM-2} = 0.0016$, $\nabla_t T^*_{FVM-3} = 0.0015$), respectively. Further analysis of gradients is detailed in the Appendix I. In this result, PINNs achieved relatively lower accuracy for dataset with greater temporal variation, as exemplified by FVM-1.

**Discussion**

This study demonstrates the viability of using PINNs to reconstruct unseen velocity and pressure fields from noisy single-color LIF temperature data and partial boundary conditions. This demonstration, conducted for highly transient mixed laminar convection flows, implies that the introduced method is applicable to a variety of flow conditions. PINNs not only filter out the high-frequency noises inherent in the single-color LIF technique, but also reconstruct the simultaneous fields of velocity and pressure without necessitating a sophisticated optical system. The reconstruction method was validated through two methods: non-simultaneous PIV and comparison with numerical simulation data. The reconstructed velocity fields were qualitatively similar to separate PIV measurements taken under identical flow conditions. Additionally, portions of numerically simulated temperature fields were used to reconstruct the complete temperature fields as well as other fields. The PINN reconstruction demonstrated $\epsilon_{T^*}$ all below 1%, while $\epsilon_{P^*}$, $\epsilon_{U^*}$ and $\epsilon_{V^*}$ were below 10%. This confirmed the PINN outputs to be reasonably accurate under the specified boundary and data constraints.

Furthermore, this study demonstrates that TL significantly reduces the computation time required for field reconstruction while maintaining accuracy for similar and unseen flow conditions. Traditional random initialization of the PINN resulted in a training duration of more than 12 hours, while the TL method reduced the training duration by a factor of 9.9. Multiple cases containing various temperature patterns were used for demonstration. The TLPINNs produced results within $\epsilon <$ 5% as compared to the traditional PINN. The best TLPINN achieved such results with only 1000 epochs, representing 1% of the epoch required for a traditional PINN.

While we have demonstrated the feasibility of PINNs for a specific type of mixed laminar convection flow, this approach has the potential to be generalized to a broader range of laminar convection flows characterized by different Re and Ri. The key factor enabling this generalization is that PINN-based flow reconstruction relies on the same governing physics laws (i.e., PDEs), across the laminar flow problems. However, when extending PINN-based reconstruction to turbulent flows, additional PDEs from turbulence models would need to be incorporated. Another extension of this research will be to address the growing interest in real-time reconstruction. We demonstrate that TLPINN can significantly decrease the computation time needed for reconstruction. However, the training still required at least 0.5 hours. By creating a library of PINNs on various flow conditions, it is possible that TL can further reduce the training time to create PINNs for near simultaneous field reconstruction.

**Methods**

**PINN.** The PINN architecture used in this work is a fully connected network comprising 10 hidden layers and each layer with 150 neurons. The select activation function is a locally adaptive sine function that incorporates a trainable neuron-level parameter [30,31]. The PINN was built using PyTorch and trained on a single Nvidia A-100 80GB GPU. For training, 100 boundary points were uniformly sampled along the $x^*$, $y^*$, and $t^*$ directions. Additionally, 20,000 collocations points were randomly sampled throughout the computation domain using the Latin-Hypercube (LHS) method [32]. The temperature data points were randomly sampled from the nodal points in either the LIF or FVM data using LHS. During training, the PDE loss was evaluated at both the temperature data points and collocation points. The optimization of the loss function employed a coupled strategy [33], involving 100,000 epochs using the Adam optimizer with a learning rate of 0.0001, followed by refinement with the L-BFGS optimizer until convergence [34,35]. On average, completing 100,000 epochs and processing 84,500 collocation points took about 12 hours and used around 25 GiB of memory. Additionally, training utilized floating-point precision with a machine epsilon of $1.19 \times 10^{-7}$.

**Transfer learning.** $PINN_0$ using 200 points per $mm^2$ and sampling period of 0.1 s is first trained for 100,000 Adam epochs and fine-tuned with L-BFGS. Transfer learning is accomplished by transferring the parameters of a trained PINN network to initialize another PINN instead of the Xavier normal scheme.

**Closed fluid loop.** A fluid channel with a rectangular cross-sectional shape consists of a heated aluminum base plate and acrylic side walls. The channel dimensions are 178 mm in length, 63 mm in width, and 20 mm in height. The heating is provided by a strip heater with a maximum power of 240 W, operating at 7.13 W/$cm^2$. The base temperature is regulated by a proportional-integral-derivative (PID) controller, which varies the heater power to maintain a constant temperature. A resistance temperature detector (RTD) located on the channel base provides the temperature input to the PID controller. To monitor temperatures, three 20 kΩ negative temperature coefficient (NTC) thermistors (±0.3℃) are positioned at the inlet, channel base, and outlet, respectively. Temperature data is recorded by a data recording system (Vernier, LabQuest Mini). The working fluid, deionized water, is degassed for 30 minutes at -84.66 kPa prior to use. The volumetric flow rate is controlled by a rotameter and a bypass tube, ensuring a constant flow rate of 3.78 liter/minute. The inlet water temperature is regulated by a separate temperature-controlled water bath and heat exchanger system.

**PIV.** Temperature and velocity visualizations are conducted using commercially available LIF and PIV systems (Dantec Dynamics, Planar LIF/PIV). The PIV system utilizes a 532 nm Nd:YAG laser operating at 10 Hz as the excitation source. Tracer particles, 20 μm diameter polyamide seeding particles (Dantec Dynamics, PSP-20), are suspended in the working fluid at a concentration of 0.25 mg/L. The fluorescent signal is captured using a charge-coupled device (CCD) camera (Dantec Dynamics, FlowSense USB 2M-165) with a manual f-stop of 2.8 and a resolution of 1920 by 1200. A synchronizer is used to coordinate the laser pulse with the camera shutter.

**Single-dye LIF.** Single color LIF measurements are performed using the same Nd:YAG laser, with Rhodamine-B (Rh-B) as the fluorescence dye. The Rh-B (Sigma-Aldrich, R6626–25 G) has an excitation peak of 546 nm and emission peak of 567 nm, and its fluorescence emission is sensitive to temperature changes. The fluorescent light passes through an optical filter (Dantec Dynamics, 9080C0561) is tuned to wavelengths greater than 532 nm. Calibration of the fluorescent intensity was conducted to establish a linear relation between intensity and temperature [36,37], which resulted in an average decline rate of approximately 2.1% per degree K. A dye concentration of 50 μg/L and a laser intensity of 20% (137 mJ, average power of 2.055 W) were used to achieve an optimal signal-to-noise ratio in the raw intensity output [3,38,39].

**FVM simulation.** A 3D finite volume model (FVM) for an open channel was developed to validate the experimental and reconstruction results using Ansys Fluent 2023 R2. The inlet boundary condition was set to a constant volumetric flow rate of 3.7 liters per minute and constant temperature of 30°C. All surfaces were treated as adiabatic except the channel base, which was subject to a uniformly distributed constant temperature of 32°C. Thermistor measurements from experiment indicate that the inlet and base temperatures were relatively constant with fluctuations within 0.2°C. The open surface on top of the channel was treated as a free-slip surface (zero shear in the parallel direction). The flow condition demonstrated in the model entailed the use of the laminar model with the energy equation. Mesh and time step independence tests were performed. The final mesh size and timestep size were, 0.5 mm and 1 ms, respectively. The simulation spanned a duration of three minutes, and temperature fields were sampled at an interval of 0.1 s ($\Delta t_{Data}$), resulting in a total dataset of 1800 frames.

**Acknowledgements**

This work was supported by two National Science Foundation grants under Grant No. 2053413 and 2337973.

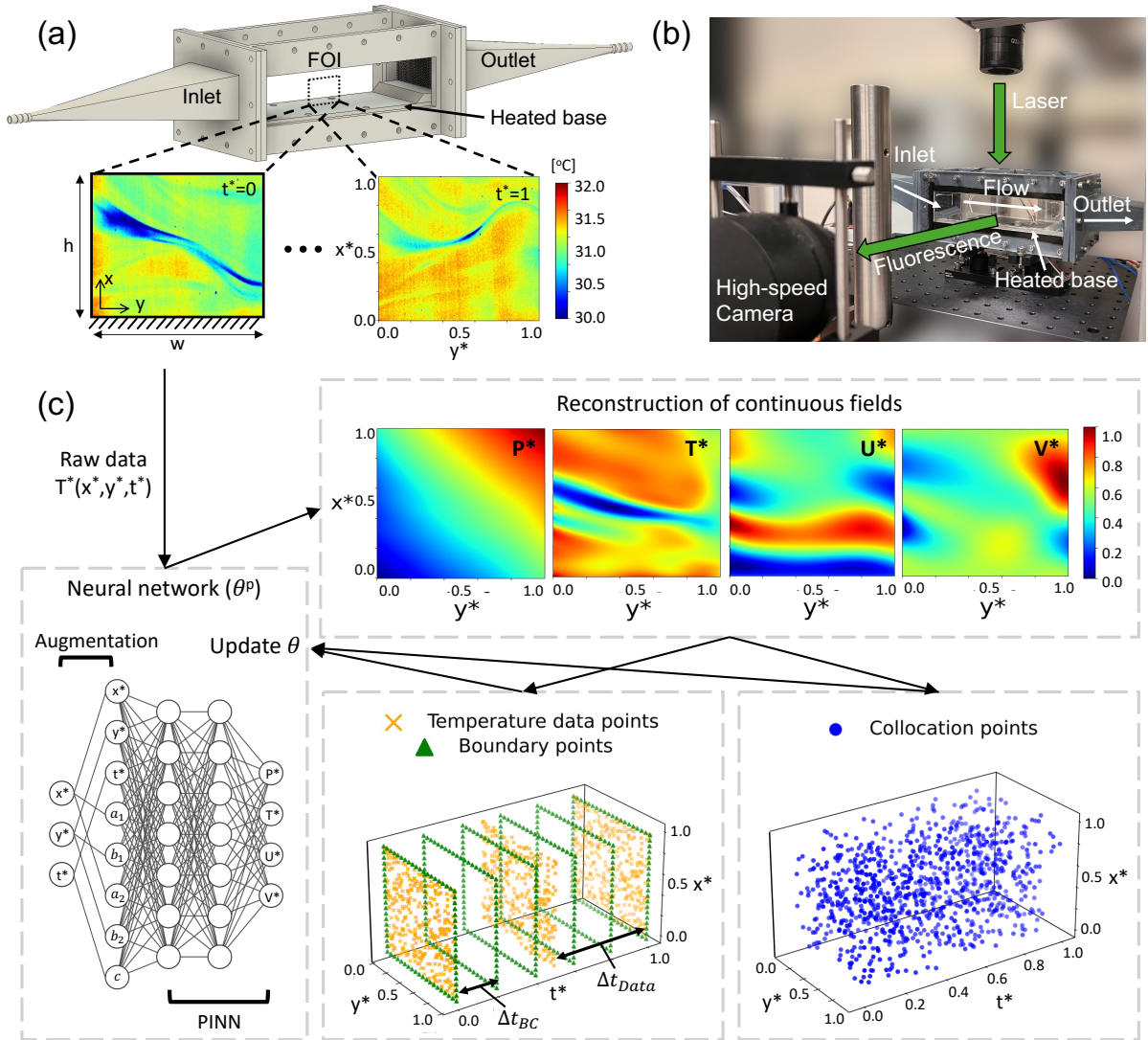

Figure 1. (a) 3D schematic of a channel with heated base and free top surface, with flow entering the inlet and exiting the outlet. 2D representation of LIF temperature sampling technique within the FOI. (b) Experimental setup showing image capturing with laser positioned on top of channel, and high-speed camera capturing the fluorescence from side of channel. (c) Overview of multiple-scale approach MSPINN for field reconstruction process. Neural network takes inputs of spatial-temporal coordinates from collocations, boundary, and LIF data points and outputs predictions for four fields (P, T, U, V). Network is updated to minimize the mean squared error (MSE).

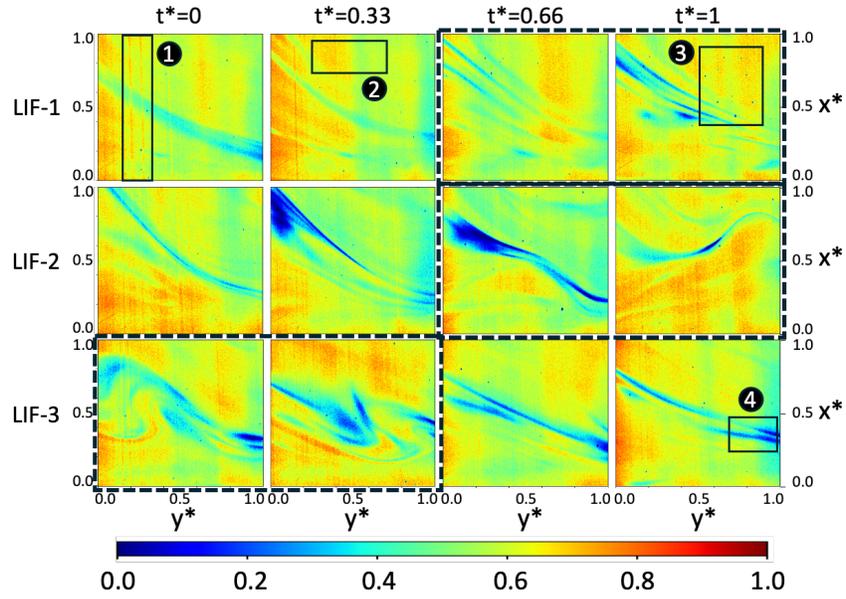

Figure 2. Three LIF datasets used for reconstruction shown at four different t*. Experimental noise in solid boxes is labeled as: 1) narrow vertical lines, 2) large-scale uneven backgrounds, with a reddish background on the left and greenish background on the right, 3) discontinuous dotted regions, and 4) discontinuous temperature fields. Dashed boxes indicate regimes of interest which include complex flow structures.

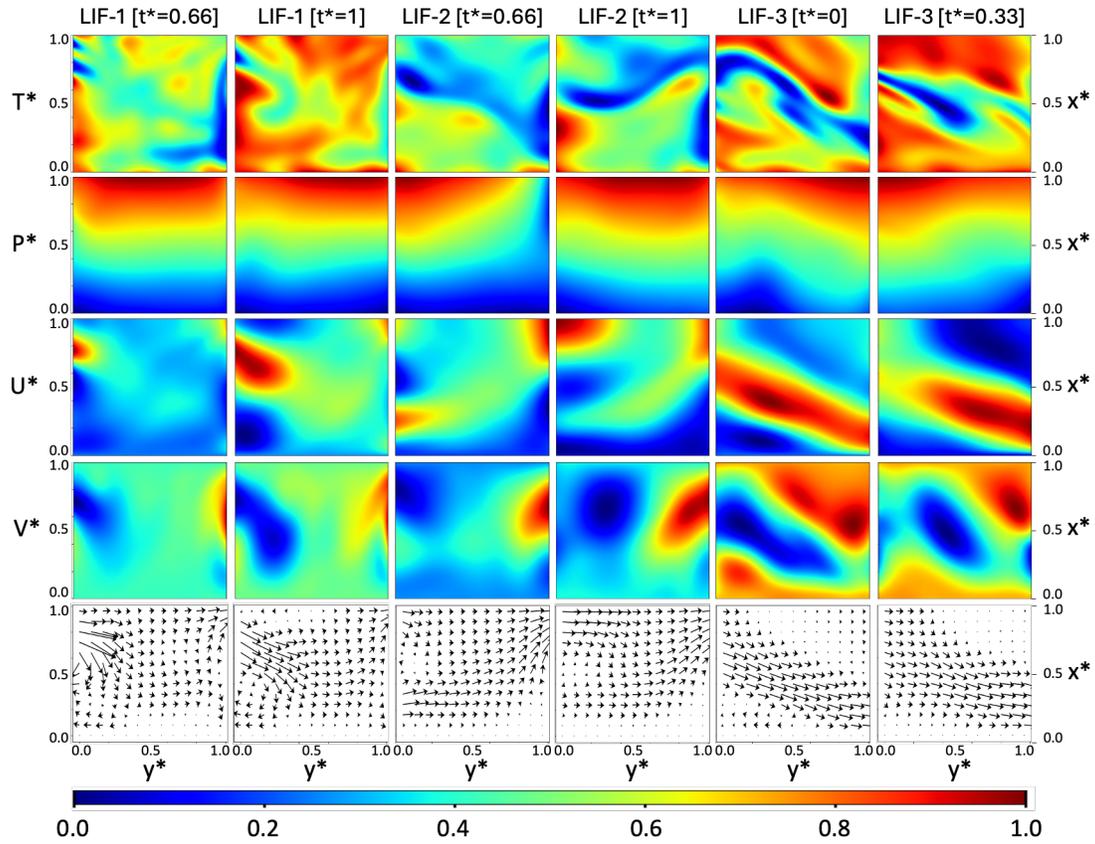

Figure 3. Reconstruction results of the three LIF datasets at timesteps that show complex flow development. In all cases, a cold stream enters from the left. Then, the cold stream gradually falls in LIF-1, gradually rises in LIF-2, and rotates slightly counterclockwise in LIF-3. The respective pressure and velocity fields along with a 2D velocity vector field are also shown.

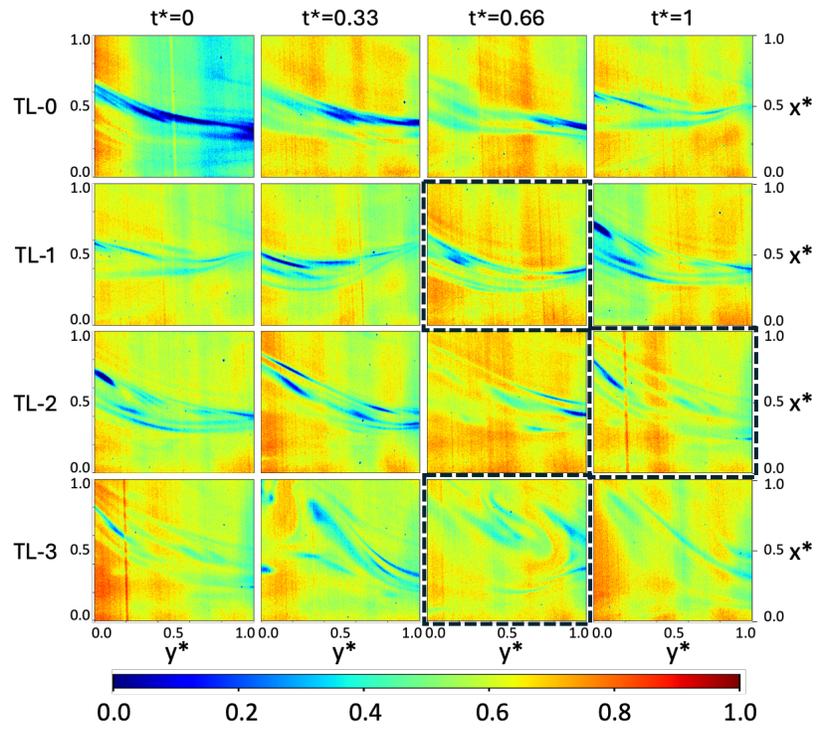

Figure 4. Four LIF datasets used for transfer learning reconstruction shown at four t*. TL-0 is used to train a baseline PINN that is used in subsequent training of TLPINNs. The datasets are sequential in which the next dataset succeeds the former (e.g., TL-1 $t^*=0$ is identical to TL-0 $t^*=1$). Dashed boxes indicate regimes of interest which include complex flow structures.

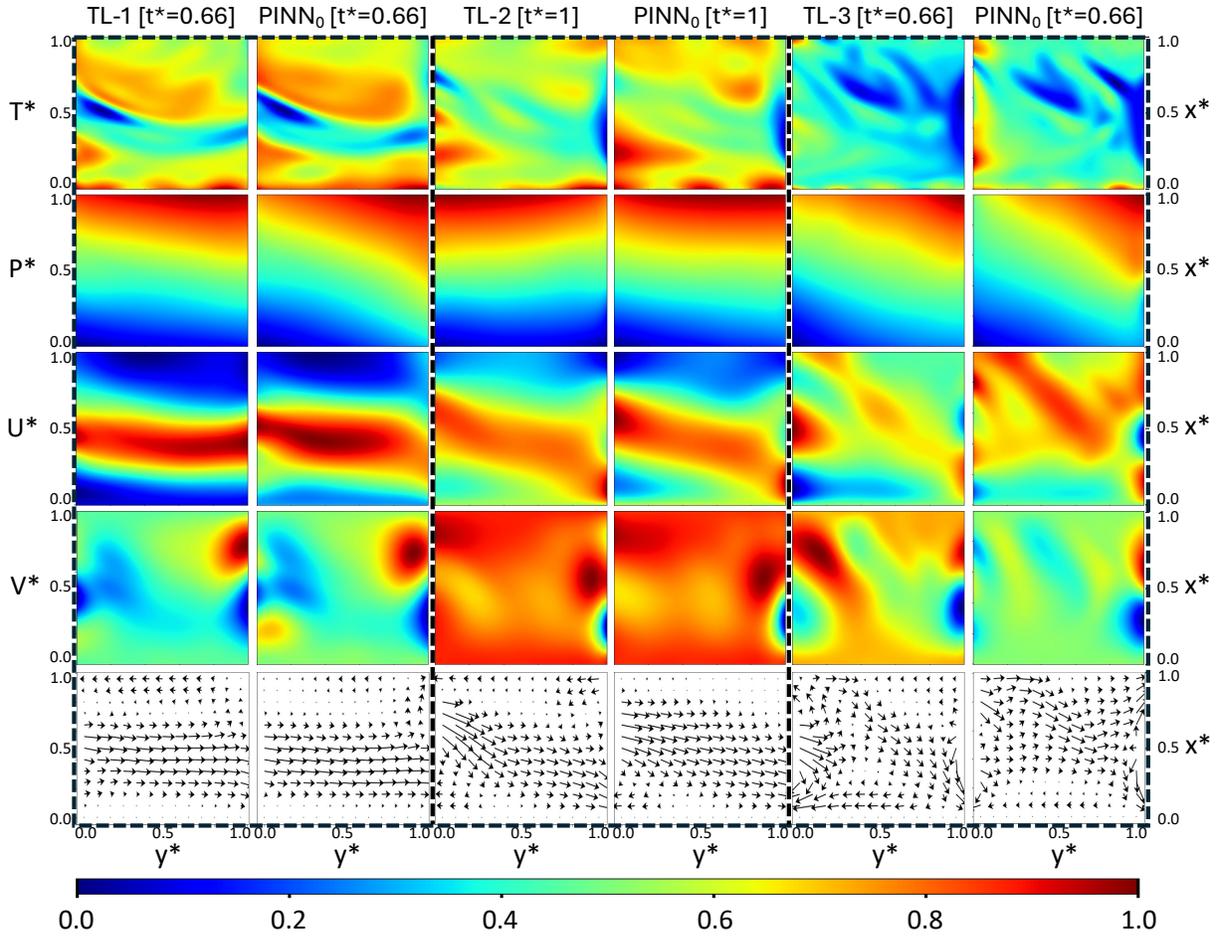

Figure 5. Reconstruction results of the three TL datasets at timesteps that show complex flow development. TLPINNs produce nearly identical reconstructions compared to the traditional PINN (PINN$_0$). TL-1 and TL-2 include a low temperature stream that enters from the right and exhibits relatively simple velocity distributions. TL-3 includes a low temperature stream that curls back onto itself that creates complex flow patterns, which is reflected in the velocity vector fields.

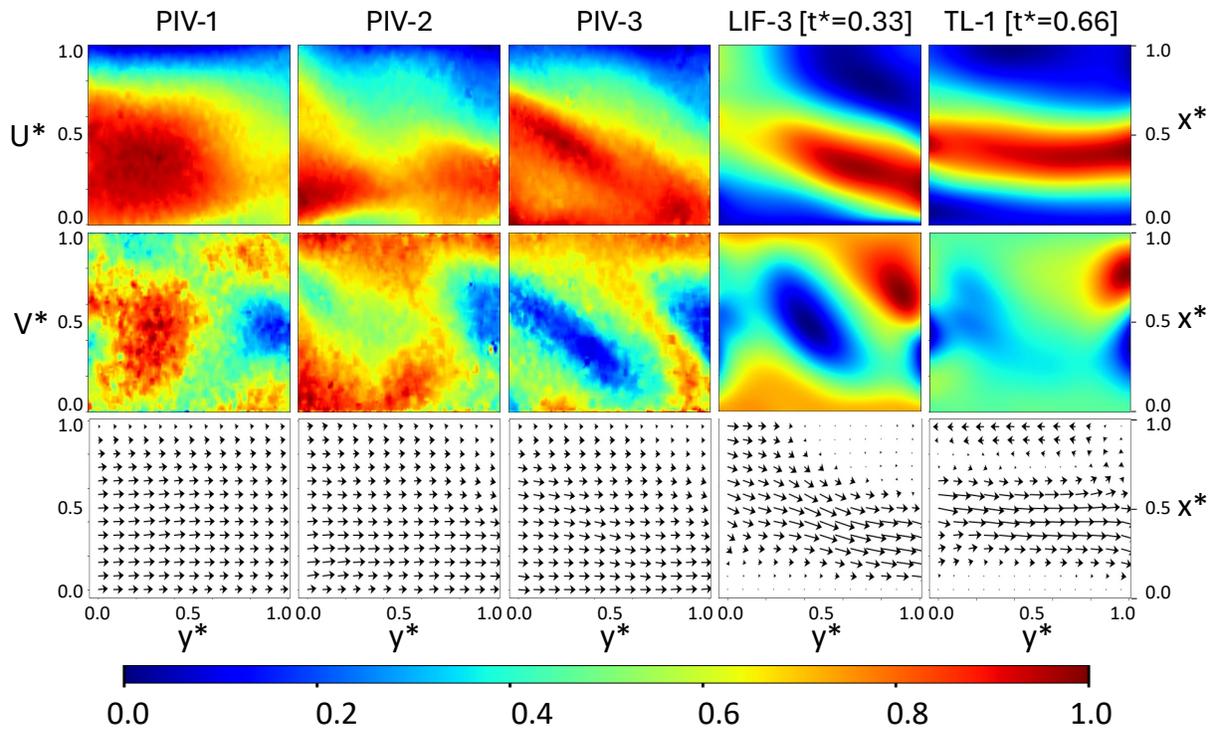

Figure 6. Three non-simultaneous PIV snapshots taken under identical flow conditions as the LIF experiment. For all snapshots, a higher $U^*$ region is present in the center with lower velocities at the top. A low $V^*$ region is present on the right side. LIF-3 and TL-1 are also shown to demonstrate similar reconstructed flow patterns.

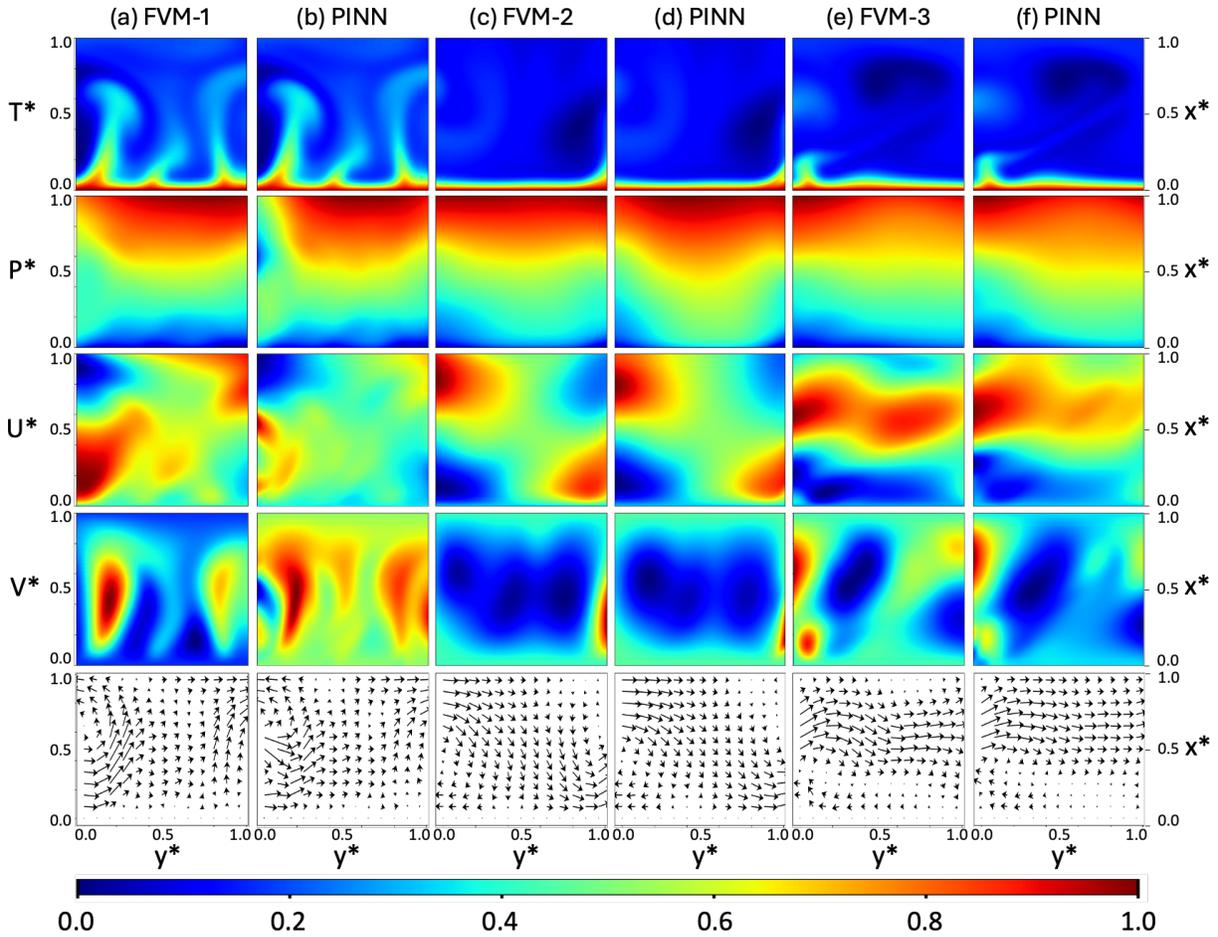

Figure 7. Reconstruction results of the three FVM datasets at timesteps that show complex flow development. PINNs under the given boundary and data constraints can reconstruct fields with a high degree of accuracy. FVM-2 and FVM-3 include a small rising plume from the base and have relatively simple velocity distributions. FVM-1 includes three plumes and has a complex temperature distribution, which necessitated a data loss weight of 50 for acceptable results.

# APPENDIX

**Table of contents**

## [A] Experiment schematic

Figure A-1. Schematic of experimental setup with fluid loop and data signal loop.

**[B] Governing Equations**

The governing equations used to constrain the PDE loss in PINNs are often non-dimensionalized and normalized to generate inputs and outputs between -1 and 1. We formulate the problem such that the inputs are between 0 and 1. Furthermore, terms within each equation should have relatively similar magnitudes to prevent stiff equations, which causes convergence issues with PINNs [24–27]. The equations used are,

$$\frac{\partial U^*}{\partial x^*} + \gamma \frac{\partial V^*}{\partial y^*} = 0 \tag{B-1}$$

$$St\frac{\partial U^*}{\partial t^*} + U^*\frac{\partial U^*}{\partial x^*} + \gamma V^*\frac{\partial U^*}{\partial y^*} + \frac{\partial P^*}{\partial x^*} - \frac{1}{Re}\left(\frac{\partial^2 U^*}{\partial x^{*2}} + \gamma^2 \frac{\partial^2 U^*}{\partial y^{*2}}\right) = 0 \tag{B-2}$$

$$St\frac{\partial V^*}{\partial t^*} + U^*\frac{\partial V^*}{\partial x^*} + \gamma V^*\frac{\partial V^*}{\partial y^*} + \gamma \frac{\partial P^*}{\partial y^*} - \frac{1}{Re}\left(\frac{\partial^2 V^*}{\partial x^{*2}} + \gamma^2 \frac{\partial^2 V^*}{\partial y^{*2}}\right) + RiT^* = 0 \tag{B-3}$$

$$St\frac{\partial T^*}{\partial t^*} + U^*\frac{\partial T^*}{\partial x^*} + \gamma V^*\frac{\partial T^*}{\partial y^*} - \frac{1}{Pe}\left(\frac{\partial^2 T^*}{\partial x^{*2}} + \gamma^2 \frac{\partial^2 T^*}{\partial y^{*2}}\right) = 0 \tag{B-4}$$

where the non-dimensional parameters are denoted with (*). Non-dimensional numbers are defined as Reynolds numbers (Re=$U_0$w/$\nu$), Peclet number (Pe=$U_0\rho_0$wc/k), aspect ratio ($\gamma$=w/h), Strouhal number (St=w/$U_0 t_{exp}$), and Richardson number (Ri=g$\beta$($T_{max}$-$T_{min}$)w/$U_0^2$). The reference values are, width of the FOI (w), height of the FOI (h), data length ($t_{exp}$), inlet temperature ($T_{min}$), and base temperature ($T_{max}$). The reference velocity ($U_0$) is chosen to yield Ri=1 and is $U_0 = \sqrt{g\beta(T_{max} - T_{min})w}$. Water properties at 30°C are density ($\rho_0$=995 kg/m$^3$), thermal conductivity (k=0.623 W/m K), thermal expansion coefficient ($\beta$=0.0003344/K), heat capacity (c=4.182 kJ/kg K), and kinematic viscosity ($\nu$=7.26E-7 m$^2$/s). The non-dimensional parameters are defined as follows, x$^*$=x/w, y$^*$=y/h, t$^*$=t/$t_{exp}$, U$^*$=U/$U_0$, V$^*$=V/$U_0$, P$^*$=P/($\rho_0 U_0^2$), and T$^*$=(T-$T_{min}$)/($T_{max}$-$T_{min}$).

## [C] MSPINN

Starting with the non-dimensionalized and normalized governing equations in Appendix B, we define an additional length ($\delta$) and time ($\tau$) scale. The rescaled variables are X=x/$\delta$, Y=y/$\delta$, and Z=t/$\tau$. Using the chain rule, we can express (B-1) through (B-4) with the rescaled variables. For example: $\frac{\partial U^*}{\partial x^*} = \frac{\partial U^*}{\partial X}\frac{\partial X}{\partial x} + \frac{\partial U^*}{\partial x}\frac{\partial x}{\partial x} = U_X^* \frac{1}{\delta} + U_x^*$, where the subscripts denote a partial derivative. Terms that appear with scales on different orders of magnitudes are examined. Because of the normalization done to arrive at (B-1) through (B-4), most terms appear with order unity, except for non-dimensional number terms: St, 1/Re, and 1/Pe. Thus, these terms should appear in the scaling and (B-1) can be ignored. Because we set Ri=1, the scaling for (B-2) and (B-3) are identical, which leaves (B-2) and (B-4) to be analyzed.

$$St(U_Z^*\frac{1}{\tau} + U_t^*) + U^*(U_X^*\frac{1}{\delta} + U_x^*) + \gamma V^*(U_Y^*\frac{1}{\delta} + U_y^*) + (P_X^*\frac{1}{\delta} + P_x^*) -$$

$$\frac{1}{Re}((U_{XX}^*\frac{1}{\delta^2} + \frac{2}{\delta}U_{Xx}^* + U_{xx}^*) + \gamma^2(U_{YY}^*\frac{1}{\delta^2} + \frac{2}{\delta}U_{Yy}^* + U_{yy}^*)) = 0 \quad \text{(C-1)}$$

$$St(T_Z^*\frac{1}{\tau} + T_t^*) + U^*(T_X^*\frac{1}{\delta} + T_x^*) + \gamma V^*(T_Y^*\frac{1}{\delta} + T_y^*) -$$

$$\frac{1}{Pe}((T_{XX}^*\frac{1}{\delta^2} + \frac{2}{\delta}T_{Xx}^* + T_{xx}^*) + \gamma^2(T_{YY}^*\frac{1}{\delta^2} + \frac{2}{\delta}T_{Yy}^* + T_{yy}^*)) = 0 \quad \text{(C-2)}$$

Taking the order of magnitudes of each term in (C-1) leads to, $\frac{St}{\tau}$, $St$, $\frac{1}{\delta}$, 1, $\frac{1}{\delta^2 Re}$, $\frac{1}{\delta Re}$, and $\frac{1}{Re}$. The $\frac{1}{\delta}$ term can be eliminated with a constant scaling to the inner solution velocities. Balancing 1 with $\frac{1}{\delta^2 Re}$ and $\frac{St}{\tau}$ leads to a dominant balance with $\delta$=Re$^{-0.5}$ and $\tau$=St. Similar analysis with (C-2) will give $\delta$=Pe$^{-0.5}$ and $\tau$=St. To augment the original input to the scaled input, we simply multiply the original input by the inverse of the scales. Details on dominant balance and multiple-scale analysis can be found in [38].

**[D] PINN loss functions**

Loss functions used in PINNs constrain the outputs to specified initial conditions, boundary conditions, data points, and governing equations. Boundary conditions used in this study are no-slip on the bottom surface and free-slip on the top surface. Data points used are temperature measurements. Governing equations are discussed in Appendix B. No initial conditions are used. Thus, the composite loss function (L) takes the form,

$$L = L_R + L_{BC} + L_D$$

Where $L_R$ is the PDE loss, $L_{BC}$ is the boundary loss, and $L_D$ is the data loss. Each loss is computed over the collocation points, boundary points, and data points, respectively. Governing equations (B-1) through (B-4) are formulated as residuals and denoted as $e_1$ through $e_4$, respectively. The losses are expressed as,

$$L_R = \sum_{i=1}^{N_R+N_D} \sum_{k=1}^{4} (|e_k(x_i, y_i, t_i)|^2)$$

$$L_{BC} = \sum_{i=1}^{N_{BC}} (|\psi(x_i, y_i, t_i) - \psi_i|^2)$$

$$L_D = \sum_{i=1}^{N_D} (|T^*(x_i, y_i, t_i) - T_i^*|^2)$$

Where $N_R$, $N_{BC}$, and $N_D$ represent the number of collocation points, boundary points, and data points, respectively. $\psi$ represents the Dirichlet boundary condition variable (e.g., $U^*$ and $V^*$). Mean square error is used as the minimization criterion.

**[E] Spatial-temporal density tuning**

The baseline PINN model and LIF-1 was used to determine the spatial and temporal density of temperature data points required for convergence. Spatial density was first examined with eight cases, 50, 100, 200, 300, 400, 500, 600, and 1000 points per mm$^2$, with temporal density at 0.1 s and over 40 frames, which is less than the total 100 frames but necessary to keep computations on a single GPU with 80 GiB of memory. Error analysis was performed over five trials against the 1000 points per mm$^2$ case as it was expected to yield the most accurate result. Each case was trained five times and the results averaged. Figure E-1 shows the mean and standard deviation of error for each spatial density, in which $\epsilon_{T^*}$ is mainly used to determine convergence since temperature data is used for training. 200 points per mm$^2$ was found to be the minimum spatial density that yielded sufficient convergence ($\Delta\epsilon_{T^*}$<1%) according to Fig. E-1 (b).

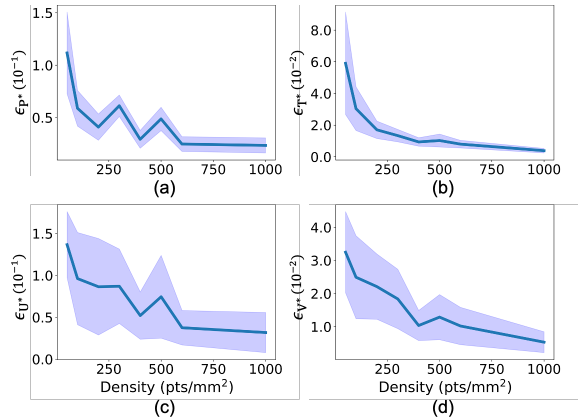

Figure E-1. Error $\varepsilon$ of PINN reconstruction at different data spatial densities using LIF-1 with mean and standard deviation. A minimum of 200 points per mm$^2$ was found to be necessary for $\Delta\epsilon_{T^*}$ <1%.

Temporal density was next examined using a spatial density of 200 points per mm$^2$. Seven cases, 0.1, 0.2, 0.3, 0.4, 0.5, 0.6, and 0.7 seconds sampling periods over all 100 frames were investigated. Error analysis over five trials was performed against the 0.1 s case as it was expected to yield the most accurate result. Each case was trained five times and the results averaged. Figure E-2 shows the mean and standard deviation of error for each temporal density. Similarly, $\epsilon_{T^*}$ was used to determine convergence, in which 0.2 s was found to be the maximum sampling period that yielded sufficient convergence ($\epsilon_{T^*}$ <10%) according to Fig. E-2 (b).

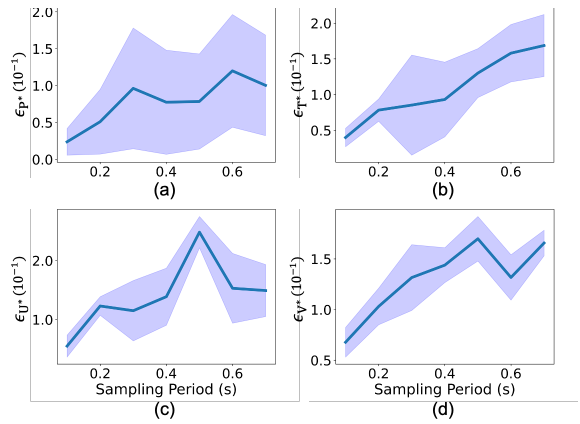

Figure E-2. Error $\varepsilon$ of PINN reconstruction at different data sampling periods using LIF-1 with mean and standard deviation. A maximum of 0.2 s was found to be necessary for $\epsilon_{T^*}$ <10%.

**[F] Loss vs epoch**

A sample of the individual training loss for LIF-PINN (TL-1, PINN$_0$), TLPINN (TL-1, PINN-2) and FVM PINN (FVM-3) were plotted to demonstrate convergence of PINN under noisy data inputs. The final PDE loss, boundary loss, and data loss were, $3.66×10^{-6}$, $1.73×10^{-4}$, and $6.84×10^{-2}$ for LIF-PINN, $7.36×10^{-6}$, $2.71×10^{-4}$, and $6.87×10^{-2}$ for TLPINN, $8.15×10^{-6}$, $2.77×10^{-5}$, and $2.97×10^{-5}$ for FVM. Each component of the composite loss is plotted against epochs in Fig. F-1.

The magnitude of data loss is significantly higher in PINNs utilizing LIF data compared to FVM (order of $10^3$). This is caused by the noisy data that fail to satisfy the PDEs, which does not allow efficient minimization of the PDE loss. PDE loss remains consistent across each PINN, indicating good convergence of LIF data with governing equations. Boundary loss for LIF-PINN and TLPINN are nearly identical, while FVM PINN is an order of magnitude smaller, which could be caused by conflicts between boundary conditions and LIF data.

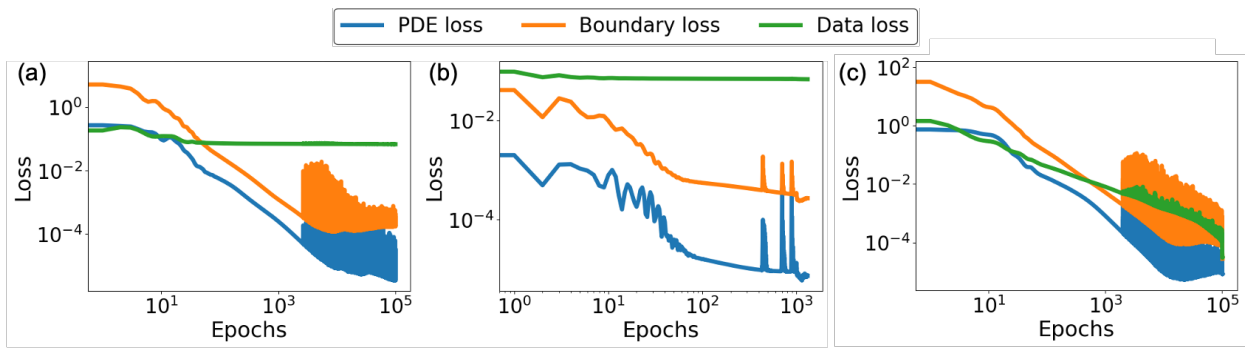

Figure F-1. Breakdown of the composite loss over the training epoch for (a) LIF-PINN, (b) TLPINN, and (c) FVM PINN. The data loss in LIF-PINN and TLPINN plateaus at a relatively large magnitude due to the presence of noise, while a clean dataset would follow a loss plot similar to (c).

**[G] TLPINN configuration study**

Prior to training different PINNs using transfer learning, a baseline PINN (PINN$_0$) needs to be established. Since the flow structures are not expected to be very different between PINNs, by allowing initialization close to the final solution, we can significantly decrease the training epochs needed for convergence. In this study, four different TL configurations are examined. All PINNs are trained with identical hyperparameters with only variations in Adam optimization epochs and are all fine-tuned until convergence with L-BFGS. The four configurations are as follows: 1) No TL at 100,000 Adam epochs, which is used to compare against the traditional PINN training method (PINN-1). 2) TL at 1000 Adam epochs (PINN-2). 3) TL at 500 Adam epochs (PINN-3). 4) TL with no Adam epochs (PINN-4). PINN$_0$ used 200 points per mm$^2$ and sampling period of 0.1 s (same spatial density and twice the temporal density compared to the PINN-1 through PINN-4). For visualization purposes we show the transfer learning results for TL-4 in Fig. G-1, for which the raw LIF image is given in Fig. 2 as LIF-1 at $t^*=0$. We note that to get comparable results, it was necessary to train PINN-1 for 200,000 Adam epochs instead of 100,000. Through visual inspection, it can be concluded that PINN-2 and PINN-3 produced the closest temperature reconstruction to PINN$_0$. Comparison of $P^*$ shows that there are minimal deviations from PINN$_0$ except for PINN-1. Lastly, $U^*$ and $V^*$ reconstructions were closest for PINN-3 and PINN-2, respectively.

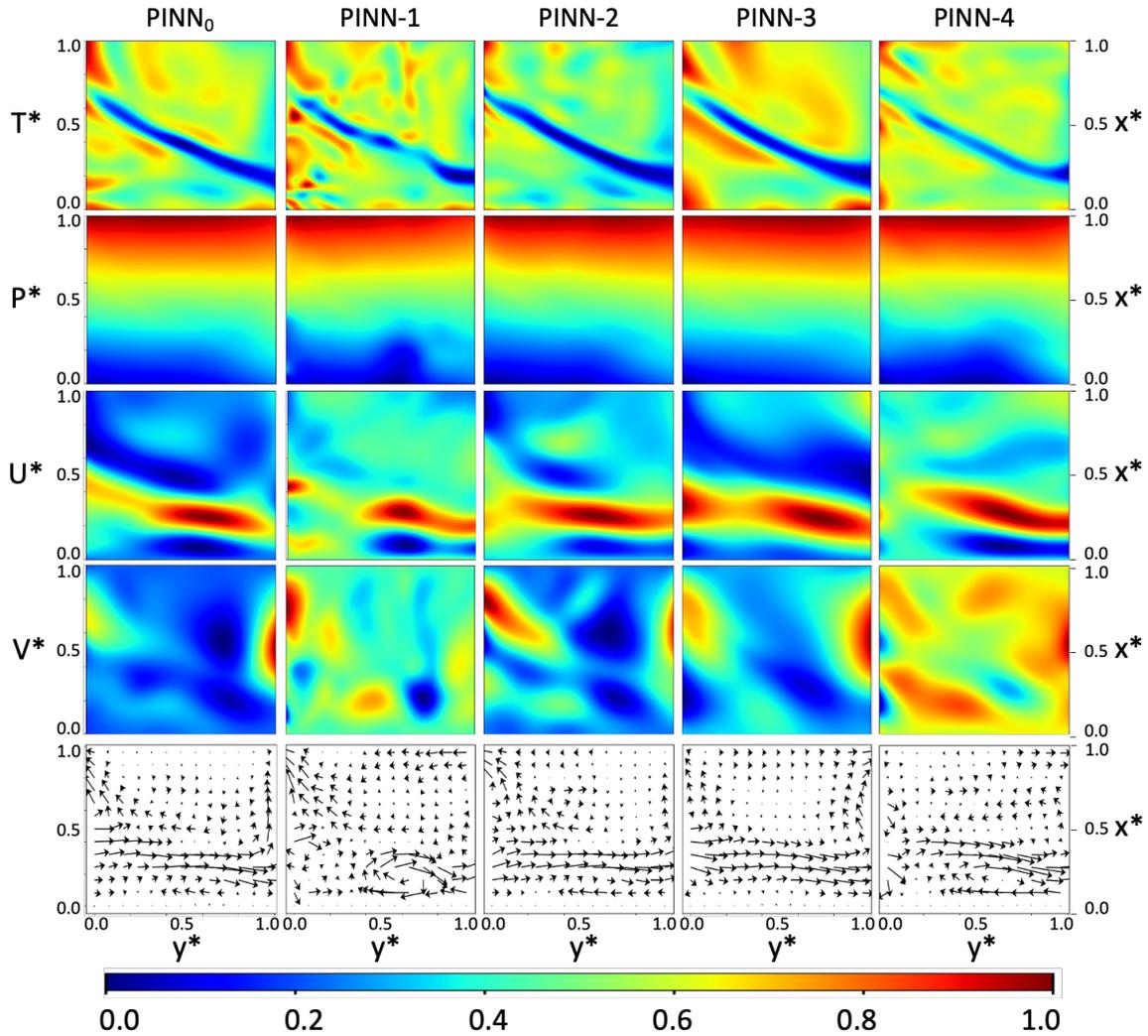

Figure G-1. Reconstruction results of different TLPINN optimization configurations. All TLPINNs have a low temperature stream entering from the left, in which visual comparison yields PINN-3 as the best temperature reconstruction. Overall, PINN-2 and PINN-3 provide the closest reconstruction results compared to PINN$_0$.

Error plots for reconstruction of PINN-1 through PINN-4 over five trials are shown in Fig. G-2. It is observed that the TLPINNs performed similarly and in some cases better than the traditional PINN-1. PINN-3 underperforms for $U^*$ in all cases, which could indicate insufficient training epochs since PINN-2 performed better. Analysis of the errors gives the conclusion that performance from best to worse are, PINN-4, PINN-1, PINN-2, and PINN-3, respectively. This leads to an interesting conclusion that direct L-BFGS optimization without any stochastics through Adam gives better transfer learning. It is evident that without tuning the Adam optimizer learning rate, the PINN is stochastically led away to another local optima.

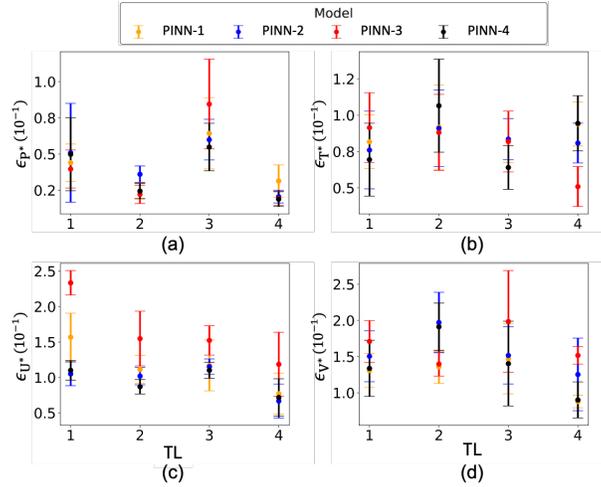

Figure G-2. Error $\varepsilon$ of TLPINN reconstruction using four different datasets and four configurations with mean and standard deviation. All average $\epsilon_{T^*}$ were below 12% and reconstruction errors for other fields were below 25%.

The computation time for each Adam epochs tested are given in Table G-1 for the four TL cases. The average compute time for each PINN configurations were, 12.45 hr, 8.32 hr, 0.85 hr, 0.75 hr, and 0.84 hr, respectively. PINN-2 and PINN-4 have similar compute time for all TL cases, while PINN-3 was slightly faster. With the error analysis and computation time, we conclude that PINN-2 and PINN-4 are the best performing TLPINNs. The traditional method of training (i.e., PINN-1) yielded similar results but with far greater training cost. Compared to PINN$_0$ and PINN-1, transfer learning was able to achieve approximately 14.8 times and 9.9 times less computational time on average, respectively.

Table G-1. Computation time for each PINN configuration when trained for each TL case given in hours.

|  | TL-1 | TL-2 | TL-3 | TL-4 |
|---|---|---|---|---|
| $PINN_0$ | 12.28 | 12.32 | 12.76 | 12.42 |
| PINN-1 | 8.28 | 8.06 | 8.42 | 8.5 |
| PINN-2 | 0.54 | 0.6 | 1.34 | 0.94 |
| PINN-3 | 0.82 | 0.76 | 0.6 | 0.8 |
| PINN-4 | 0.64 | 0.64 | 1.16 | 0.92 |

## [H] PIV Flow Development

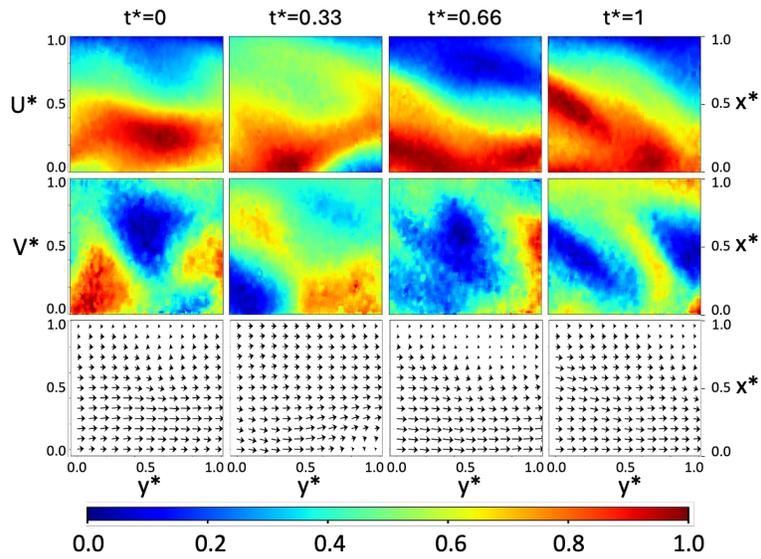

Figure H-1. Velocity development of PIV data under identical flow conditions as LIF. A high $U^*$ stream is visible in the center of the channel and is surrounded by slower fluid on the top and bottom. Some oscillation of flow is visible in the velocity vector field.

**[I] Error Analysis**

FVM data was used to analyze the best-case scenario for reconstruction. Different channel sections were examined to explore the relationship between data acquisition position and reconstruction error. Each PINN was trained on a single section over a range of 100 frames. Figure I-1 shows $\epsilon$ combined from multiple individual PINNs for each field when analyzed across 300 frames, in which later frames generally had higher $\epsilon$ and channel section did not impact $\epsilon$. On average, $\varepsilon_{T^*}$ remains below 2%, which indicates good reconstruction of temperature. However, this does not correlate with the ability of PINNs to reconstruct other fields, which indicates that other metrics should also be examined. The temporal gradients in Fig. I-2 were analyzed to determine the impact of vanishing gradients. Overall, it is shown that a smaller gradient usually leads to higher $\epsilon$. Since reconstruction is done from the temperature fields, relative magnitude of $\nabla_t T^*$ gives a good estimation on the expected reconstruction error, which propagates to other fields as well.

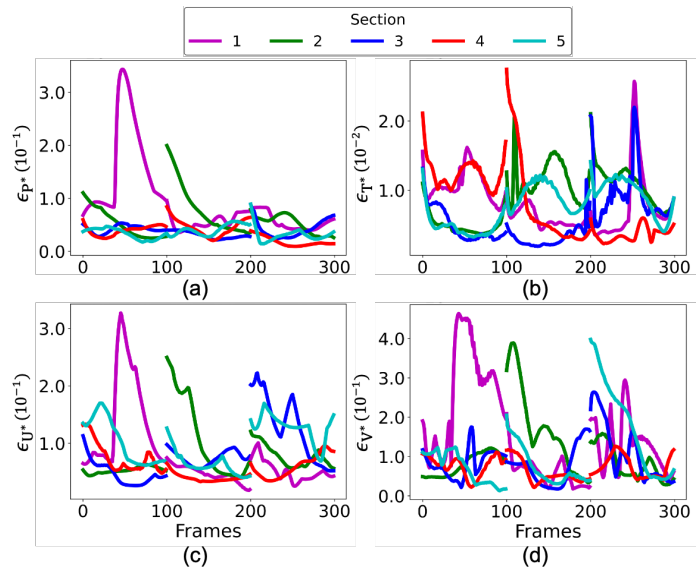

Figure I-1. $\varepsilon$ of PINN reconstruction using FVM data across a 300-frame segment and at various channel sections. Although the average $\varepsilon_{T^*}$ for section 1 frame 0-100 is below 2%, it is still higher than the other sections, which corresponds to outlier $\varepsilon$ for other fields.

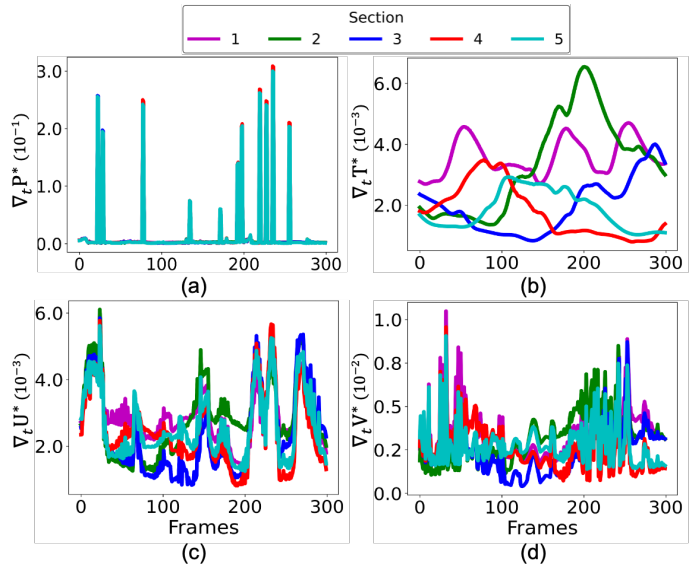

Figure I-2. $|\nabla_t F^*|$ (temporal gradient) for each field F, where F refers to field variables P, T, U, V. Spikes in $|\nabla_t P^*|$ indicate large changes in pressure between adjacent frames, but do not seem to correlate with error.